# LIGHT FIDELITY (LI-FI) BASED INDOOR COMMUNICATION SYSTEM

Farooq Aftab, Muhammad Nafees Ulfat khan,Shahzad Ali

School of Computer and Communication Engineering,University of Science and Technology Beijing china (USTB)

*ABSTRACT*

*Indoor wireless communication is an essential part of next generation wireless communication system.For an indoor communication number of users and their device are increasing very rapidly so as a result capacity of frequency spectrum to accommodate further users in future is limited and also it would be difficult for service providers to provide more user reliable and high speed communication so this short come can be solve in future by using Li-Fi based indoor communication system. Li-Fi which is an emerging branch of optical wireless communication can be useful in future as a replacement and backup of Wireless Fidelity (Wi-Fi)for indoor communication because it can provide high data rate of transmission along with high capacity to utilize more users as its spectrum bandwidth is much broader than the radio spectrum. In this paper we will look at the different aspects of the Li-Fi based indoor communication system,summarizes some of the research conducted so far andwe will also proposed a Li-Fi based communication model keeping in mind coverage area for multiple user and evaluate its performance under different scenarios .*

*KEYWORDS*

*LightFidelity,Indoor communication, Visible light communications, Wi-Fi.*

## 1. INTRODUCTION

Light Fidelity (Li-Fi) is a new paradigm of revolution which is a continuation of the trend to move toward higher frequency spectrum[2] in the field of indoor wireless communication. If we talk about high speed indoor wireless communication, Li-Fi can bring new dimension in term of data communication speed by utilizing visible light spectrum. The concept behind this technology is that the data can be transmitted with the help of light emitting diode (LED) bulbs and transmission rate can be control by using intensity of LED bulb which can be varies even faster than light intensity human eye can observe [1]. Keeping in mind number of users increase day by day and heavy traffic of data, Li-Fi technology can be used as a solution to provide users an environment of high speed data transmission.

Li-Fi system is a bi-directional multiuser communication system and it could be classified as nanometer-wave communication system [2]. Li-Fi based communication system is different from Visible light communication (VLC) system[2] because VLC is only a point to point communication system while Li-Fi is a proper wireless based networking system which supports point to multipoint communication.In Li-Fi system, data rate can be related with LEDs so LEDs selection plays a vital role. Parameters like LEDs size, ON-OFF speed and number of LEDs can affect data rate of the communication model.  Data rate is inversely proportional to the Size of LEDs which means smaller the size of LED more will be the data rate. If ON-OFF speed of LED is faster than we can transmit data at higher rates in the form of 1's and 0's. Higher the number of LED's in a system results in more transmission of data. In Li-Fi based system LED Panel (LP) is a Light source which can perform the function of illumination and data communication at a same time. Depending upon the situation and requirement of indoor communication model a Single LED bulb can also act as a LP and it is also possible that multiple LEDs bulb are combined together to form a single LP . Number of LEDs in a single LP depends upon the size of room and number of users to accommodate at a specific time. LED Placement approach (LPA) plays an important role in Li-Fi based indoor communication system because it can limit data rate and affect the communication. The placement of LP in indoor environment must be adjusted in such a way that every Li-Fi user can achieve high intensity of light.

Significant research effort is being directed toward the development of Li-Fi indoor system. For an indoor system the LED is declared as most power efficient illuminating device for future indoor lighting system [3]. In [4] researchers investigate how the distance between LEDs can change the behavior of indoor communication system. The impact of multipath reflections on a two-dimensional in door positioning is investigated in [5] .In current indoor visible light positioning systems, several algorithms are proposed to calculate the receiver position as in [6] researchers propose a novel architecture system that can be used for both indoor positioning and communications. In [7] light positioning architecture for a typical room has been investigated by considering different performance related parameters. In [8] researcher designed an optimal constellation for Indoor 2×2 MIMO based Communication system under arbitrary channel correlation. Keeping in mind all the research conducted by different researchers until now this Paper is organized as follow In Section II, we have discussed theoverview of Li-Fi based indoor communication system which is consists of detail regarding key elements that Indoor communication system must possess in order to satisfy the demand of implementation of Wireless communication system. In Section IIIwe have described Scenario based indoor Li-Fi architecture design in which we have proposeddifferent types of LP in term of coverage area and Section IV is consists of performance evaluation of proposed designed system followed by conclusion.

## 2. LI-FI BASED INDOOR COMMUNICATION SYSTEM OVERVIEW

### 2.1. LI-FI TRANSMITTER AND RECEIVER

The transmitter in an indoor Li-Fi system is an LED bulb. The most likely candidate for front-end transmitter devices is incoherent solid-state lighting LEDs due to their low cost. Due to the physical properties of these components, information can only be encoded by using the intensity of the emitted light. Different LED's of different color like red, blue, orange, yellow can be used in Li-Fi Communication System. But if we talk about high data rates, 1 Giga bits per second has been reported using phosphor-coated white LEDs [9] and 3.4 Giga bits per second has been red-green-blue (RGB) LEDs [10], the highest speed that has ever been reported from a single color incoherent LEDs is 3.5 Giga bits per second.

LED luminaire commonly use white light to perform both the function of illumination and communication. One way of producing white light is to use blue LED with yellow phosphor coating. When a beam of blue light passes through yellow phosphor coating layer it becomes white light. Another way is to use a combination of red, green and blue (RGB) LEDs .when red, green and blue light properly mixed together it becomes white light.As the light emitted by LEDs are incoherent in nature so therefore there is a need of Intensity Modulation (IM). In IM signal is modulated in to optical signal of instantaneous power. This signal is received at a receiver by using Direct Detection (DD) method. In Direct Detection (DD) a photodiode is used to convert the optical signal power into a proportional current.

Table 1. Method to generate white light.

| White light production method | Advantage | Disadvantage |
| --- | --- | --- |
| Blue light with yellow phosphor coating | Easy to implement and cost effective | phosphor coating limits the speed at which LED can switched to a few MHz |
| RGB light | Easy to modulate the data using three different color wavelength LEDs | Not cost effective |

As Li-Fi system is based on IM/DD therefore Avalanche Photo Detector (APD) is more effective as compare to PIN based PD.

## 2.2. MODULATION

In Li-Fi based system, Dimming based modulation schemes are most commonly used modulation schemes which are single carrier based schemes. In dimming based modulation schemes desire data rate is achieve by controlling the On-Off level of LED. On-off keying (OOK),Pulse Width Modulation (PWM),Pulse position modulation (PPM), Variable pulse position modulation (VPPM),Overlapping PPM (OPPM) and optical spatial modulation (OSM) are the main dimming based modulation schemes which can be implemented in Li-Fi based indoor system .Dimming based modulation schemes are explained in table 2.

To achieve higher data rate and to decrease the effect of distortion and interference, multicarrier modulation can also be useful in Li-Fi based communication system but multicarrier modulation schemes are less energy efficient. One of the most common schemes is OFDM [11] but OFDM based signal is complex and bipolar in nature so to implement OFDM for Li-Fi system some modifications are required in conventional technique for better performance. In [12] researchers proposed a Asymmetrically-Clipped Optical OFDM (ACO-OFDM) in which odd subcarriers are modulated DC-biased Optical OFDM (DCO-OFDM) [13] is a scheme in which all subcarriers are modulated and unipolar signal is generated by adding positive direct current. ACO-OFDM is more energy-efficient as compare to DCO-OFDM. The relationship between light emitted by LED and current is nonlinear so this nonlinearity based nature of LED affects the performance of OFDM based modulation schemes.

There are some modulation schemes which are designed to support both purpose of communication and illumination by using multicolored LEDs. Color shift keying (CSK) is a scheme [14]in which signals are encoded into color intensities emitted by red, green and blue (RGB) LEDs .The constant color is maintained by mapping the transmitting bits in to instantaneous chromatics of LEDs to ensure constant luminous flux. CSK has Reliability on LED performance due to constant luminous flux and has no flicker effect over all frequencies. In [15] researchers proposed a Metameric modulation (MM) which modulate data in the visible spectrum while maintaining a constant lighting state.MM has a better Color quality control and higher energy efficiency. Color intensity modulation (CIM) proposed in [16] provides dimming in color space. CIM also satisfy the need of color matching and increases the data rate in signal space for multicolored LED based system.

Table 2. Dimming based modulation schemes.

| Modulation scheme | working | Advantage | Disadvantage |
| --- | --- | --- | --- |
| On-off keying(OOK) | OOK is a dimming based modulation scheme which transmits data by sequentially turning the LED on and off. | OOK provides a good trade-off between system performance and implementation complexity | Increasing or decreasing the brightness of the LED from 50% dimming level would cause the data rate to decrease. |
| Pulse Width Modulation (PWM) | In PMW signal pulses carry the modulated signal in the form of a square wave and the widths of the pulses are adjusted based on the desired level of dimming | PMW achieves the dimming without changing the intensity level of pulses. | Data rate is low. |
| Pulse position modulation (PPM) | PPM allows one pulse per symbol duration .The symbol duration is divided into *time* slots of equal duration and a pulse is transmitted in one of the *time* slots. | PPM is more power-efficient as compare to OOK. | PPM has a lower spectral efficiency and limited data rate. |
| Variable pulse position modulation (VPPM) | VPPM is a combination of PPM and PMW .In VPPM width of signal pulses is changed according to a specified brightness level to support Dimming. | Better spectral efficiency as compare to PPM. | VPPM lack power efficiency. |
| Overlapping PPM (OPPM) | OPPM allows more than one pulse to be transmitted during the symbol duration and the different pulse symbols can be overlap. | OPPM can achieve a higher spectral efficiency as compared to PPM and VPPM. | wide range of dimming levels can be obtained due to overlapping. |
| optical spatial modulation(OSM) | OSM is a single-carrier transmission modulation technique in which input bits sequence is mapped and correspond to a certain LED index. | Power and bandwidth efficient. | performance can be deteriorates due to the high optical channel correlation because it depend upon transmitter and receiver units |

## 2.3. COMMUNICATION MODES IN LI-FI BASED SYSTEM

One possible mode of communication in a Li-Fi system can be a Bi-directional based Indoor Communication System .If we considered a case where two users want to exchange their data using an indoor Li-fi based system then there is a need of duplexer on both sides. Duplexer is used here to support bi-directional communication by separating the transmitted and receive signal. When user1 wants to send data it will pass through the LED driving circuit toward the LED Light source panel.LED light source panel is consists of LED bulbs while LED driving circuit is used to power LED bulb. This circuit must provide sufficient current to light the LED at the required brightness, but must limit the current to prevent damaging the LED. Receiver can receive an optical signal with the help of photo detector and then data will be passing through the trans-impedance amplifier and duplexer. Then eventually receiver will receive its data as shown in figure 1.

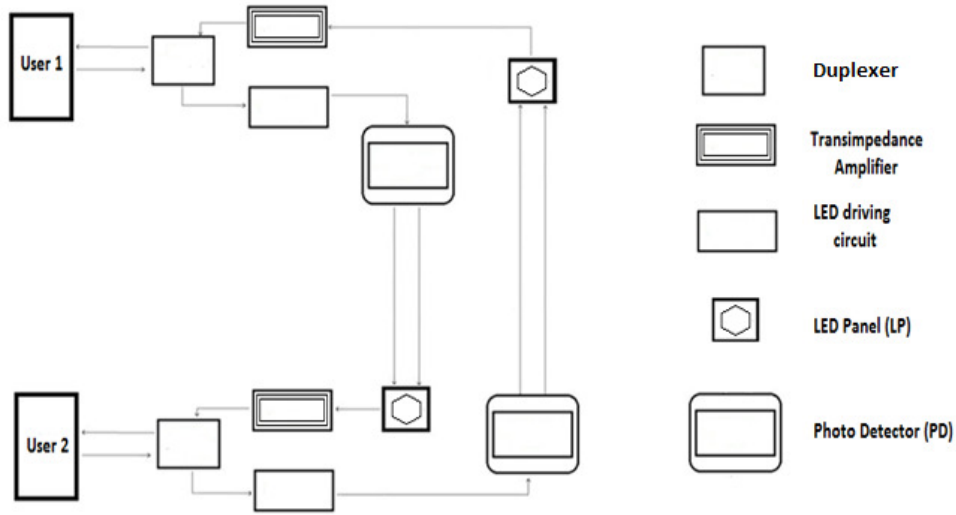

Figure 1. Bi-directional based Architecture of Indoor Communication System

If we consider multiple input and multiple output (MIMO) based Indoor Communication system then signal processing play a vital role to ensure desire data rates in the presence of high Signal to noise ratio (SNR).

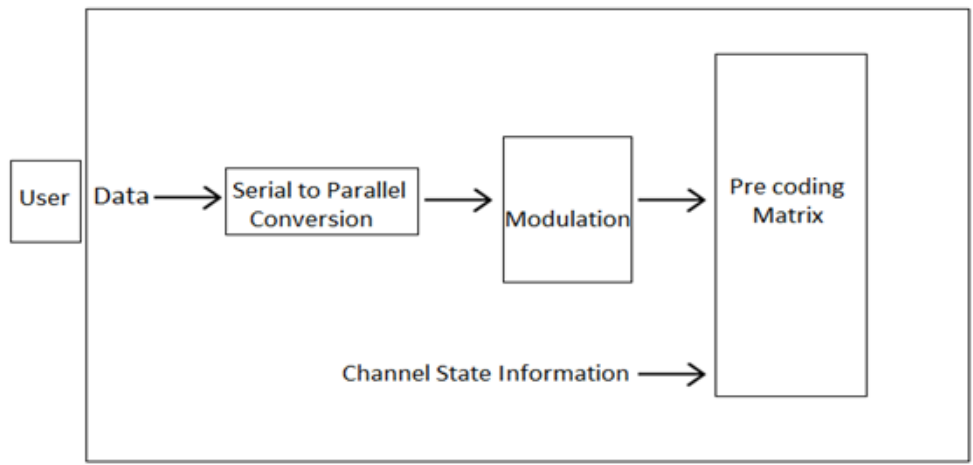

Figure 2. Data Processing Unit of User Terminal

In an indoor system, MIMO technique can be useful to have parallel data transmission to increase the data rate [17] But the MIMO channel need to be highly correlated, which increases the complexity of decoding parallel channels in the receiver [18] We consider here as shown in figure 3.

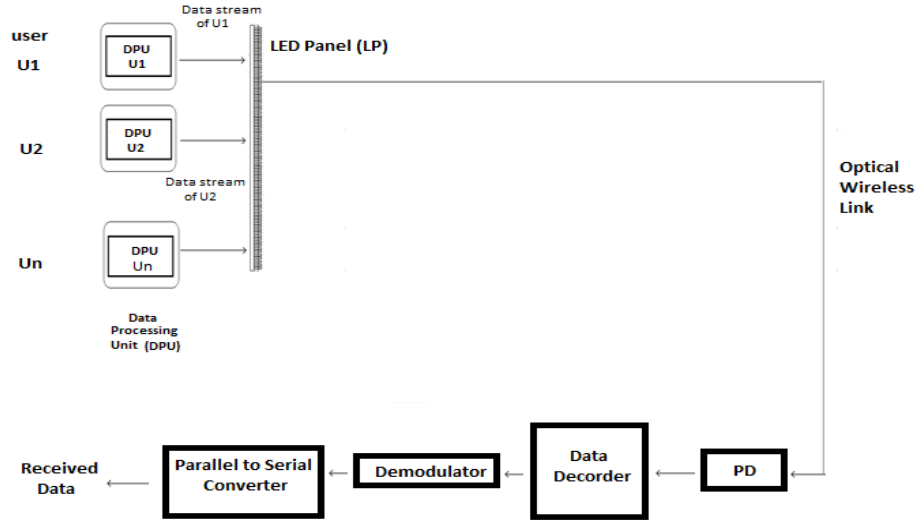

Figure 3. Architecture of MIMO based Indoor Communication System

At transmitter end we can avoid the complications of multiple user interface (MUI) by processing data signal at each terminal using Data Processing nodes. Data Processing Unit as shown in figure 3 is consists of Modulation and pre-coding Technique. This can allow us easy demodulation of signal at receiver end without any complication of high processing and will reduce burden on user terminals in result we can also achieve low energy consumption.

## 3. SCENARIO BASED INDOOR LI-FI ARCHITECTURE DESIGN

For an indoor communication system coverage area is important therefore we have proposed a scenario based Li-Fi system design in which coverage area is taken in term of LP. We can control the coverage area by considering fixed, moveable and hybrid types of LP.

### 3.1. FIXED LP

To consider the importance of coverage area in a fixed LP case we have consider here a scenario in which we have 2 users one is using laptop while other is using tablet personal computer .If we suppose in our Li-Fi system 1 LED is acting as a LP then covering area of LP can be adjusted in two ways dedicated LP approach and single LP with wide coverage area as shown in figure 4.

In section A we have placed a fixed LP of wider coverage area to accommodate two users. U1 and U2 are both connected to T1 because they both lie inside the coverage area of single LED. In section B we used a dedicated approach in which for each user individual LP is assigned.

International Journal of Computer Networks & Communications (IJCNC) Vol.8, No.3, May 2016

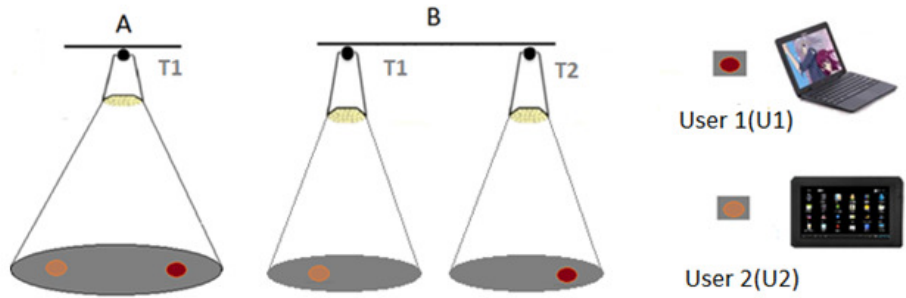

Figure 4. LP in term of Covering Area

### 3.2. MOVEABLE LP

The moveable LP is an approach in which LED bulb can rotate at a certain angle to adjust the coverage area according to the requirement of receiver placement. As shown in figure 5 A coverage area of a transmitter is controlled by rotation of LED at certain angel. As angle of irradianceis the angle with respect to the transmitter perpendicular axis and angle of incidence is the angle with respect to the receiver axis. By using moveable LP approach the light intensity on PD can be controlled by using certain angle of irradiance as a result we can also control the distance between the receiver and the transmitter to achieve desire normalized received power.

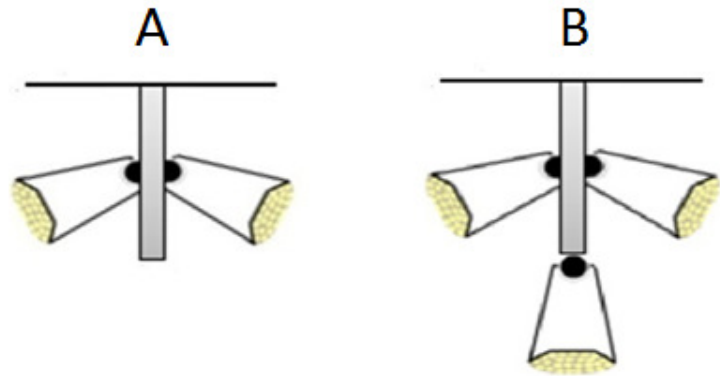

Figure 5. Moveable LP approach

### 3.3. HYBRID LP

Hybrid approach can also be used by deploying both moveable and fixed LP at a same time as shown in figure 5 B and 6.

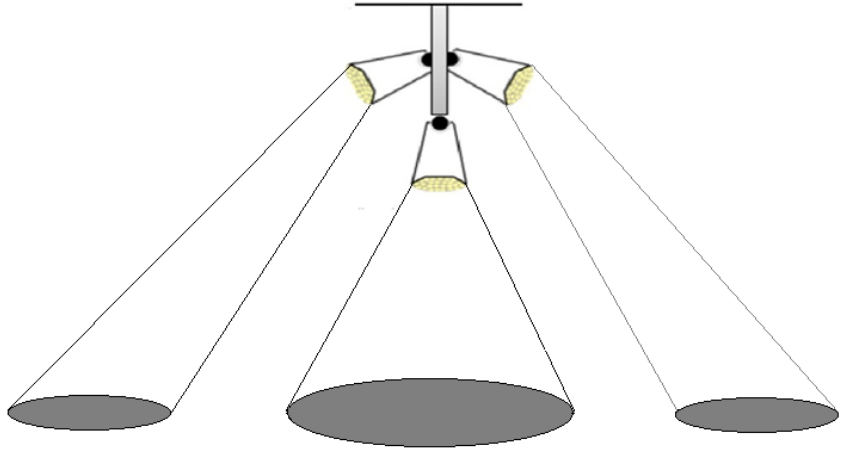

Figure 6. hybrid LP approach

## 4. PERFORMANCE EVALUATION OF PROPOSED DESIGN

For performance evaluation we assumed line of sight(LOS) link between transmitter and receiver. The LEDs are assumed to be lambertian sources and receiver is fixed at a certain height. The angle of irradiance is denoted by $\theta$ and angle of incidence is denoted by $\varphi$.

For moveable LP case we assume the receiver is fixed at a place so we have a fixed angle of incidence $\varphi = 45^o$ and we can adjust transmitter according to the placement of receiver.

SNR can be expressed as the ratio of the received visible light power and ambient noise as shown below [19]

$$\text{SNR} = \frac{\gamma^2 P^2_{rec}}{\sigma^2_{total}} \tag{1}$$

Where $P_{rec}$ the received signal is power and $\sigma_{total}$ is the total noise variance whereas $\gamma$ represent brightness level. Table 3 and figure 7 shows corresponding SNR value with respect to transmission angle to achieve BER of $10^{-5}$ for a PD receiver.

Table 3. Performance evaluation based on SNR for PD

| Angle of irradiance( $\theta$ ) | Angle of incidence($\varphi$) | Signal to noise ratio(SNR) |
|---|---|---|
| $65^o$ | $45^o$ | 128dB |
| $68^o$ | $45^o$ | 124dB |
| $70^o$ | $45^o$ | 121dB |
| $75^o$ | $45^o$ | 117dB |
| $78^o$ | $45^o$ | 114dB |

In our selected scenario we can achieve desire BER value when angle of irradiance is smaller then $70^o$ while keeping angle of incidence constant for each case .when angle of irradiance is greater then $70^o$ we cannot achieve satisfactory BER which can affect the data rate and performance of system.

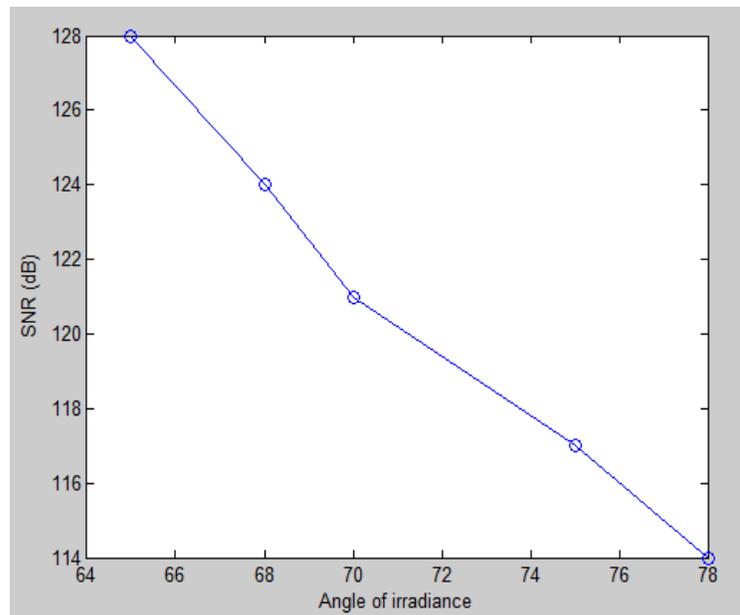

Figure 7. Performance evaluation by taking SNR with respect to angle of irradiance ($\theta$)

LP of wider coverage area which covers multiple users is better approach if we keep in mind cost efficiency but the small distance separation between the LED and PD can affect signal to noise ratio (SNR). With dedicated LP Approach we can achieve more secure, high data rate and fast communication because coverage area formed by adjacent LED is not overlapped so in this case there is a tradeoff between cost and high speed.

Moveable LP approach is also useful for practical implementation in indoor communication system because in moveable LP approach we can adjust the transmission angles of the LEDs. By transmitting a beam of light at certain angle of irradiance we can achieve desire angle of incidence as a result we can get efficient communication without overlapping coverage area between adjacent LEDs. But due to wide Field of view (FOV) of receiver the reflection components can cause the signal spread and channel bandwidth will also be decrease as a result it will limit overall data rate of system. So in moveable LP approach there is a need of tunable or fixed FOV to overcome signal spread to achieve high data rates.

Hybrid approach can also be utilized if we need to accommodate more number of users at a time and it is an efficient option of implementation without any adjacent overlapping coverage area.

## 5. CONCLUSION

We have proposed a scenario based Li-Fi system design in which coverage area is taken in term of LP. We can control the coverage area by considering fixed, moveable and hybrid types of LP. We have explained which type of LP is better in which scenario and also evaluated the effect of proposed design in term of SNR. We have also presented an overview of Li-Fi based indoor communication system. Li-Fi based indoor communication network can provide us more efficient and genuine substitute of RF based indoor wireless network and this technology has the ability to turn every light Bulb in to a Wireless Hotspot.Li-Fi based Indoor communication system has high Initial Installation cost but when it is implemented at large scale area it can accommodate us by its less operating cost like electricity bills, less operational staff and limited maintenance charges as compare to RF system. Li-Fi communication user always need line of sight connectivity with its light source therefore some advance research work is required to overcome this limitation to implement this technology in practical use. Service Providers while providing Li-Fi Indoor services has to consider major issues like reliability and availability of system and companies also need to consider how to maintain network for better performance.

## AUTHORS

**Farooq Aftab** received his BS Telecommunication Engineering degree in 2013 from Foundation University, Islamabad, Pakistan. Currently he is pursuing master degree in Information and Communication Engineering from University of Science and Technology Beijing, China. His research area is Mobile adhoc network (Manets), Light fidelity (Li-Fi) and Network coding.

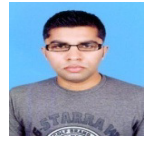

**Muhammad Nafees Ulfat khan** received his degree ofBachelor of Science in Electronics and Communication in 2013 from University of Lahore, Pakistan. Currently he is pursuing master degree in Information and Communication Engineering from University of Science and Technology Beijing, China. His research area is Light fidelity (Li-Fi) and visible light communication (VLC).

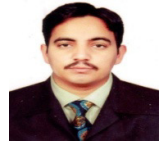

**Shahzad Ali** received his degree of M.Sc computer Science in 2013 from Abdul wali khan Mardan, Pakistan. Currently he is pursuing master degree in Information and Communication Engineering from University of Science and Technology Beijing, China. His research area is Wireless Communication.

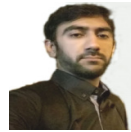